\documentstyle[preprint,aps]{revtex}

\begin{document}
\title{Vanishing Gamow-Teller Transition Rate for $A=14$ and the
Nucleon-Nucleon Interaction in the Medium}
\author{M.S. Fayache$^1$, L. Zamick$^2$, H. M\"{u}ther$^3$\\
(1) D\'{e}partement de Physique, Facult\'{e} des Sciences de Tunis\\
Tunis 1060, Tunisia\\
\noindent (2) Department of Physics and Astronomy, Rutgers University\\
Piscataway, New Jersey 08855\\
(3) Institut f\"{u}r Theoretische Physik der Universit\"{a}t
T\"{u}bingen, D-7400 T\"{u}bingen, Germany}
\date{\today}
\maketitle

\begin{abstract}
The problem of the near vanishing of the Gamow-Teller transition ($GT$)
in the $A=14$ system between the lowest $J=0^+~ T=1$ and $J=1^+~ T=0$ states 
is revisited. The model space is extended from the valence space 
$(p^{-2})$ to the valence space plus all 2$\hbar \omega$ excitations. 
The question is addressed as to what features of the effective
nucleon-nucleon interaction in the medium are required to obtain the
vanishing $GT$ strength in this extended space. It turns out that a
combination of a realistic strength of the tensor force combined with
a spin-orbit interaction which is enhanced as compared to the free
interaction yields a vanishing $GT$ strength. Such an interaction can
be derived from a microscopic meson exchange potential if the
enhancement of the small component of the Dirac spinors for the
nucleons is taken into account. 
\end{abstract}

\bigskip\bigskip\noindent 
In this work, we reconsider the old problem of the near-vanishing of the 
Gamow-Teller transition matrix element $B(GT)$ in the $A=14$ system between 
the $J=0^+~T=1$ ground state of $^{14}O$ or $^{14}C$ and the $J=1^+~T=0$ 
ground state of $^{14}N$. This is an  
allowed transition, but somehow the configurations of the initial and final 
states conspire to make this matrix element nearly vanish. Therefore the
calculation of this transition strength can serve as a very sensitive test for
the nucleon-nucleon ($NN$) interaction in the nuclear medium.

The simplest shell model configuration  consists of two holes in the $p$ shell 
for both the initial and final states. Using an $LS$ representation, the 
wavefunctions can be written as :

\[\Psi(J=0^+,T=1)  = C_i^S |^1S_0\rangle + C_i^P |^3P_0\rangle \]

\[\Psi(J=1^+,T=0)  =  C_f^S |^3S_1\rangle + C_f^P |^1P_1\rangle + 
C_f^D |^3D_1\rangle \]

It was shown analytically by Inglis \cite{inglis} that it was not possible to 
get $B(GT)$ to vanish if the two-body interaction consisted of only a central 
and a spin-orbit interaction. Afterwards Jancovici and Talmi 
\cite{talmi} demonstrated that one could get $B(GT)$ to vanish if one also 
included a two-body tensor interaction.  

What happens when we increase the model space by allowing 2$\hbar \omega$ 
configurations? Can we then get $B(GT)$ to vanish {\em without} a tensor 
interaction? We have previously performed such larger-space calculations 
\cite{fay1,fay2} but we have not specifically addressed this problem. We 
used a $G$-matrix derived from the realistic interaction Nijm II 
\cite{nijm,zheng1} which, of course, contains a tensor interaction. The 
specific result in \cite{fay1} was that in the small space ($p^{-2}$) the 
value of $B[GT(0^+~1 \rightarrow 1^+~0)]$ was 3.967, and that in the large 
space it was found to be 1.795. This is far from zero, but it is encouraging 
that higher shell admixtures will reduce $B(GT)$. We will come back to this 
later. 

Zheng and Zamick \cite{zheng2} studied the effects of 
varying the strengths of the spin-orbit and tensor interactions on $B(GT)$ in 
the small space $p^{-2}$. Indeed, the main purpose of this work will be to 
extend this study to the large space. The motivation is the following: The
interaction to be used in a rather small model space contains large
correction terms to account for the renormalization of the $NN$ interaction to
this small model space. On the other hand, the $NN$ interaction in larger 
model spaces requires renormalization only with respect to short-range 
correlations and therefore the $G$-matrix might be an appropriate
approximation.

In order to explore the sensitivity of the $GT$ strength on the spin-orbit and
tensor interactions, we employ the two-body interaction introduced in
\cite{zheng2}

\begin{equation}
V(r)=V_c(r)+x \cdot V_{s.o.}+y \cdot V_t\, ,
\label{eq:xyint}
\end{equation}

\noindent where $s.o.$ stands for the two-body spin-orbit interaction, $t$
for the tensor interaction, and $V_c(r)$ is a spin-dependent central 
interaction. Note that this interaction is not only used for the residual
interaction of the nucleons in the model space but has also been {\em
{employed to determine the single-particle part of the Hamiltonian
which is due to the interaction with the respective core.}}
The parameters $x$ and $y$ were introduced so one could easily 
vary the strengths of the spin-orbit and tensor interactions. Roughly 
speaking, $x=1,~y=1$ gives the best fit to a realistic $G$-matrix. However, 
for ($x=1,~y=1$) the value of $B(GT)$ was too large: $B(GT)=2.980$. This is 
similar to what happened with the realistic Nijm II interaction
mentioned above \cite{nijm}. It was noticed by Zheng $et. al.$
\cite{zheng2} that one could get  $B(GT)$ to vanish in at least two
ways: one way is to keep the tensor strength fixed at $y=1$ and
increase the spin-orbit strength parameter from $x=1$ to
$x=1.4$. Another way was to keep $x=1$ and decrease the tensor
strength by a factor of two ($y=0.5$). All this is in the small space.

As a first step we use the interaction of Zheng $et. al.$ \cite{zheng2} given 
in Eq. (1)  in a large space ($p^{-2}+2\hbar \omega$). To see if we can get  
$B(GT)$ to vanish without any tensor force, we set $y=0$ and vary $x$. 
The results are shown in Fig. (1) where $B(GT)$ is 
plotted $vs.$ $x$, the strength of the spin-orbit interaction. Starting from 
$x=0$, we do indeed see a rapid drop in $B(GT)$ as $x$ is increased. However, 
the curve flattens out at around $x=1.5$, and the value of $B(GT)$ is close to 
unity up to $x=7.5$. Hence it appears that, in our parameterization,
one cannot get $B(GT)$ to vanish in the large space $p^{-2}+2\hbar
\omega$ without a tensor interaction.

In Fig. (2), we show the {\em amplitude} $A(GT)$ for $x=1$ as a function of 
$y$ in the small space ($p^{-2}$). We note that the amplitude (and hence 
$B(GT)$) does go to zero, however it does so not at $y=1$ but rather close 
to $y=0.5$, about half the full tensor strength. Thus, this figure confirms 
the early work of Jancovici and Talmi \cite{talmi} that with a tensor 
interaction we can get $B(GT)$ to vanish. There is concern, however, that the 
strength of the tensor interaction needed in this small model space is
quenched by a factor of 1.33 or so as compared to the realistic estimate. 

In Fig. (3), we repeat the calculations in a large space ($p^{-2}+
2\hbar \omega$ excitations). We keep the spin-orbit strength fixed at $x=1$, 
and we vary $y$, plotting the amplitude $A(GT)$ as a function of $y$. We see 
that Fig. (3) is completely different from Fig. (2). The amplitude never 
changes sign, and hence $B(GT)$ never goes to zero. The curve is relatively 
flat from $y=0$ to $y=1.5$. Does this mean that the old ideas are wrong and 
that one cannot get $B(GT)$ to go to zero even with a tensor interaction? 

Before we jump to such a conclusion, let us repeat the calculation but with a 
stronger spin-orbit interaction. Now we keep $x$ fixed at $1.5$ rather than 1, 
and we calculate $A(GT)$ as a function of $y$. The results of these 
large-space calculations, which are shown in Fig. (4), are qualitatively 
similar to the small-space results for $x=1$. The amplitude does change sign, 
and $B(GT)$ vanishes near $y=0.75$. With a larger spin-orbit interaction, we 
regain in a large space the results that were previously obtained in a small 
space with the `free' spin-orbit interaction. The tensor interaction strength 
$y$ in the large space calculation is closer to the free-space value. All of 
this may seem somehow ad-hoc, but as we shall see next, it fits in
well with modern ideas about medium modifications of the $NN$
interaction inside the nucleus.

Relativistic mean-field studies within the framework of the so-called
quantum-hadro-dynamics or `Dirac Phenomenology' of Serot and Walecka 
\cite{serot} demonstrated that the structure of the nucleon
self-energy leads to an enhancement of the small component of the
Dirac spinors for the nucleons inside the nuclear medium as compared
to the Dirac spinors for the free nucleon. This enhancement can be
parameterized in terms of an effective Dirac mass $m_D$ for the
nucleon. The enhancement of the small component corresponds to a
reduction of the Dirac mass $m_D$ as compared to the free nucleon mass
$m$. This reduced Dirac mass yields an enhancement of the spin-orbit
splitting in the single-particle spectrum.

It was shown by Zheng $et. al.$ \cite{zzm1,zzm2} that the Dirac
Phenomenology yields non-negligible effects in nuclear structure
calculations. They demonstrated in particular that the enhancement of
the spin-orbit splitting just discussed is reproduced in nuclear
structure calculations using realistic $NN$ interactions. The
experimental data for the spin-orbit splitting of one-hole states are
only reproduced if the reduction of Dirac mass $m_D$ predicted in 
Dirac-Brueckner-Hartree-Fock calculations \cite{muth} is taken into
account in calculating the $G$-matrix elements of the 
One-Boson-Exchange interaction ($OBE$). If this relativistic feature is
ignored, the spin-orbit splitting comes out too small. It should be
noted that whereas in the relativistic Hartree-Fock method of
ref. \cite{serot} there are no pions in the theory, this is not the
case in the $OBE$ $G$-matrix calculations of ref. \cite{muth}.

The shell-model calculations of $B(GT)$ with Nijm II
\cite{nijm,zheng1} did not include this relativistic feature. 
Therefore, in the context of the ($x,y$) interaction, increasing the
spin-orbit term by putting $x=1.5$ can be interpreted as a way to
simulate the relativistic enhancement of the spin-orbit effects. Since
the spin-orbit interaction is inversely proportional to $m_D/m$ a
choice of $m_D/m=2/3$, which is a rather realistic one, would be
sufficient to increase the spin-orbit interaction by 50\% (from $x=1$
to $x=1.5$). 

Whether one should use a weaker tensor interaction inside a nucleus is more 
controversial. G.E. Brown and M. Rho \cite{brown} argue that inside a nucleus 
the masses of all mesons except the pion are less than in free space. Thus, 
the exchange of a $\rho$ meson between two nucleons would lead to a 
longer-range repulsion, thus canceling more of the attractive contribution 
due to one-pion exchange. This would yield a net weaker tensor interaction. 
However, some of the present authors \cite{fay1,fay2} have proposed an 
alternate picture of why the tensor interaction appears to be weaker inside 
a nucleus relative to free space. They call this the `self weakening' 
mechanism. Basically, the idea is that if one introduces higher-shell 
admixtures perturbatively into valence-space calculations, this will make the 
tensor interaction appear to be weaker. In the latter picture, the tensor 
anomaly is explained by doing better nuclear structure calculations. 

Of course, the two mechanisms are not mutually exclusive. In the present 
calculation, the self-weakening mechanism manifests itself in the fact that, 
in the $A=14$ beta decay, we need $y \simeq 0.5$ in the small space, but when 
higher-shell admixtures are introduced we find that $y \simeq 0.75$. 

In Table I, we depart from our phenomenological ($x,y$) interaction and show 
results with the relativistic Bonn A $G$-matrix elements. We consider the 
cases where $m_D/m$ is equal to 1, 0.75 and 0.60, and we perform the 
calculations in the small and large spaces. In this table, the $B(GT)$'s are 
shown alongside with the energy of the lowest $(J=0^+~T=1)$ state in $^{14}C$ 
relative to that of the ground state $(J=1^+~T=0)$ of $^{14}N$.

With $m_D/m=1$, we get very close to the non-relativistic matrix elements. The 
results for $B(GT)$ are very far from zero, consistent with what we obtained 
with the $(x,y)$ interaction with $x=1,~y=1$ as well as with the previously-
mentioned Nijm II interaction. In the small space, we get $B(GT)=5.294$, and 
in the large space 2.335, but at least we get closer to zero in the large 
space. 

As we decrease the Dirac effective mass, we get results closer and closer to 
zero. Finally, for $m_D/m=0.6$, we get $B(GT)=0.098$ in the small space and 
0.0018 in the large space. This is gratifying. The main reason for this 
success is of course that by decreasing $m_D/m$ we increase the spin-orbit 
splitting. Furthermore the Bonn A potential seems to be very appropriate since
it contains a tensor force which is weaker than in other realistic
$NN$ interactions\cite{rupr}. It should be reemphasized that in this
work, all the single-particle energies are calculated with 
the {\em same} interaction that is used between the valence particles
or holes. We feel this is the only way one can truly test the
correctness of a given interaction or the $G$-matrix derived
therefrom.  

It is amusing to note that in order to get $B(GT)$ to vanish for
$A=14$ one has to bring out all the artillery. When we allow
higher-shell admixtures, we are at first dismayed that -as shown in
Fig. (3)- we cannot get $B(GT)$ to vanish no matter what the strength
of the tensor interaction is. However, if we increase the strength of
the spin-orbit interaction, the ideas of Inglis \cite{inglis} and
Jancovici and Talmi \cite{talmi} are revalidated. Furthermore,
justification for this phenomenological step is afforded by the more
fundamental Dirac Phenomenology approach of Serot and Walecka
\cite{serot} and M\"{u}ther, Machleidt and Brockman \cite{muth}.

Support from the U.S. Department of energy DE-FG 02-95ER-40940 is greatly 
appreciated. M.S. Fayache gratefully acknowledges travel support from
the Universit\'{e} de Tunis, and is thankful for the kind hospitality 
of Pr. L. Zamick's Nuclear Theory Group at Rutgers University.

\samepage

\begin{table}
\caption{$B(GT)$ for $^{14}C~(J=0^+~T=1)~\rightarrow ^{14}N~(J=1^+~T=0$) with 
the Bonn A interaction}
\begin{tabular}{cccc}
$Space$ & $m_D/m$ & $E~(MeV)$ & $B(GT)$\\
\tableline
$0~\hbar \omega$ & 1    & 1.701 & 5.294\\
                 &      &       &\\
$0~\hbar \omega$ & 0.75 & 1.045 & 0.1530\\
                 &      &       &\\
$0~\hbar \omega$ & 0.60 & 1.172 & 0.0978\\
                 &      &       &\\
\tableline
                 &      &       &\\
$(0+2)~\hbar \omega$ & 1    & 1.825 & 2.335\\
                 &      &       &\\
$(0+2)~\hbar \omega$ & 0.75 & 1.426 & 0.1275\\
                 &      &       &\\
$(0+2)~\hbar \omega$ & 0.60 & 1.316 & 0.0018\\
\end{tabular}
\end{table}

\bigskip\bigskip\noindent {\bf Figure Captions}\\

{\bf Fig. (1)}: The Gamow-Teller amplitude $A(GT)$ calculated with the 
($x,y$) interaction of Eq. (1) in large ($i.e.$ $(0+2)~\hbar \omega$)
space, as a function of $x$ (the spin-orbit strength) and with $y=0$
(no tensor interaction).\\

{\bf Fig. (2)}: $A(GT)$ calculated with a spin-orbit strength of $x=1$
(free-space value), as a function of the tensor strength $y$ in small
space (0 $\hbar \omega$).\\

{\bf Fig. (3)}: Same as Fig. (2) except here the calculation is
done in large space ($(0+2)~\hbar \omega$).\\

{\bf Fig. (4)}: $A(GT)$ calculated with an enhanced spin-orbit strength
of $x=1.5$, as a function of the tensor strength $y$ in large 
space ((0+2) $\hbar \omega$).

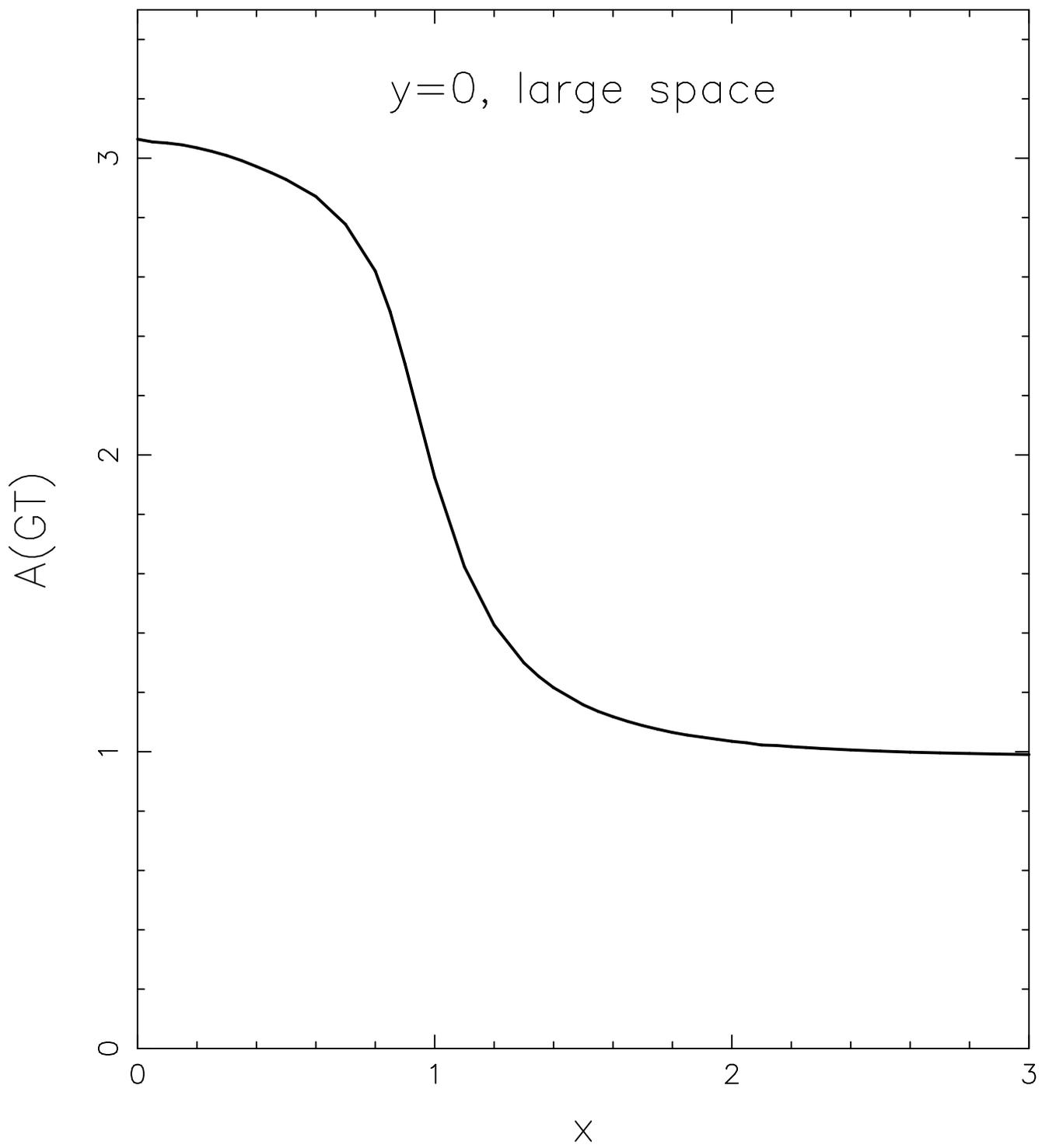
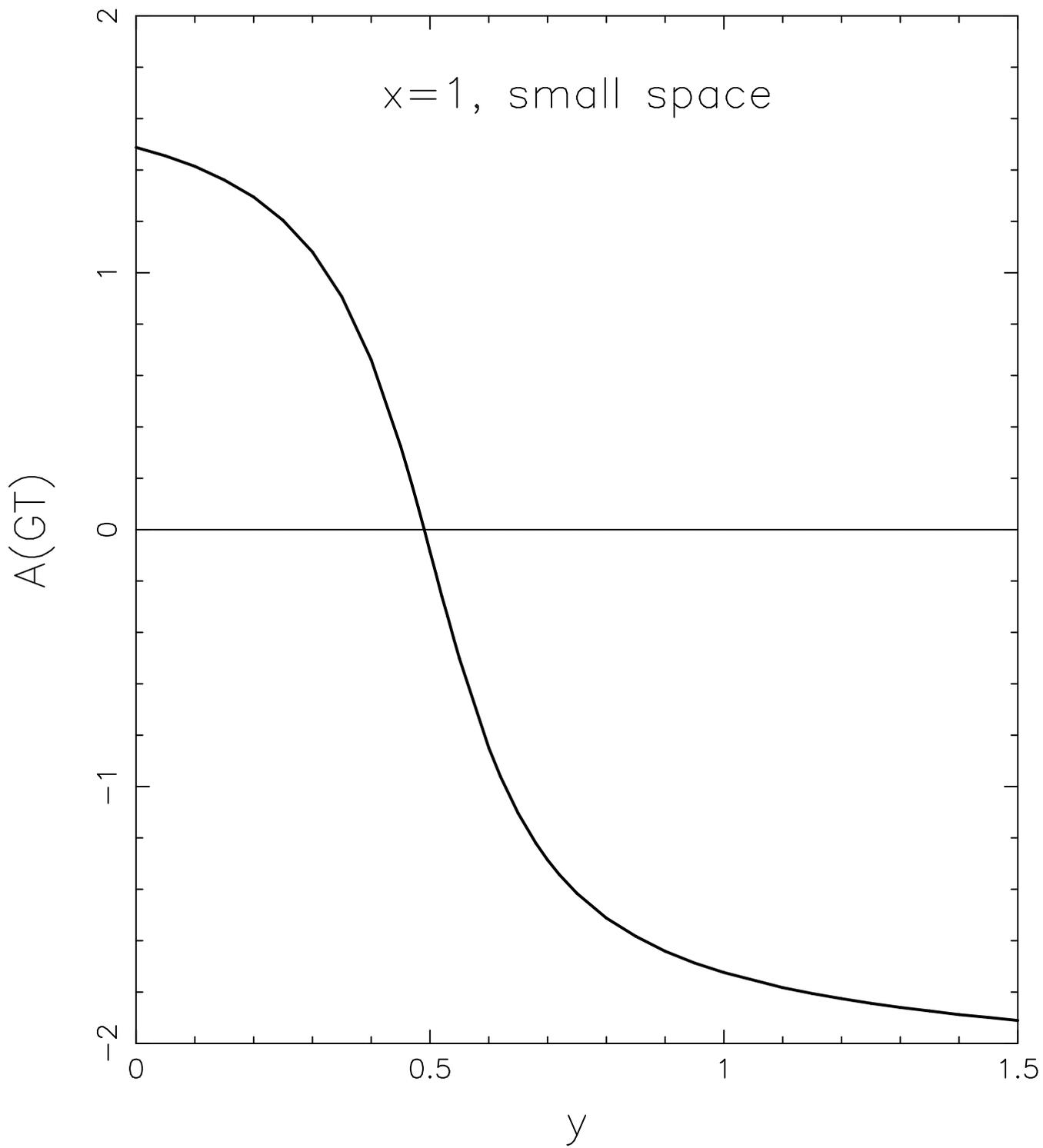
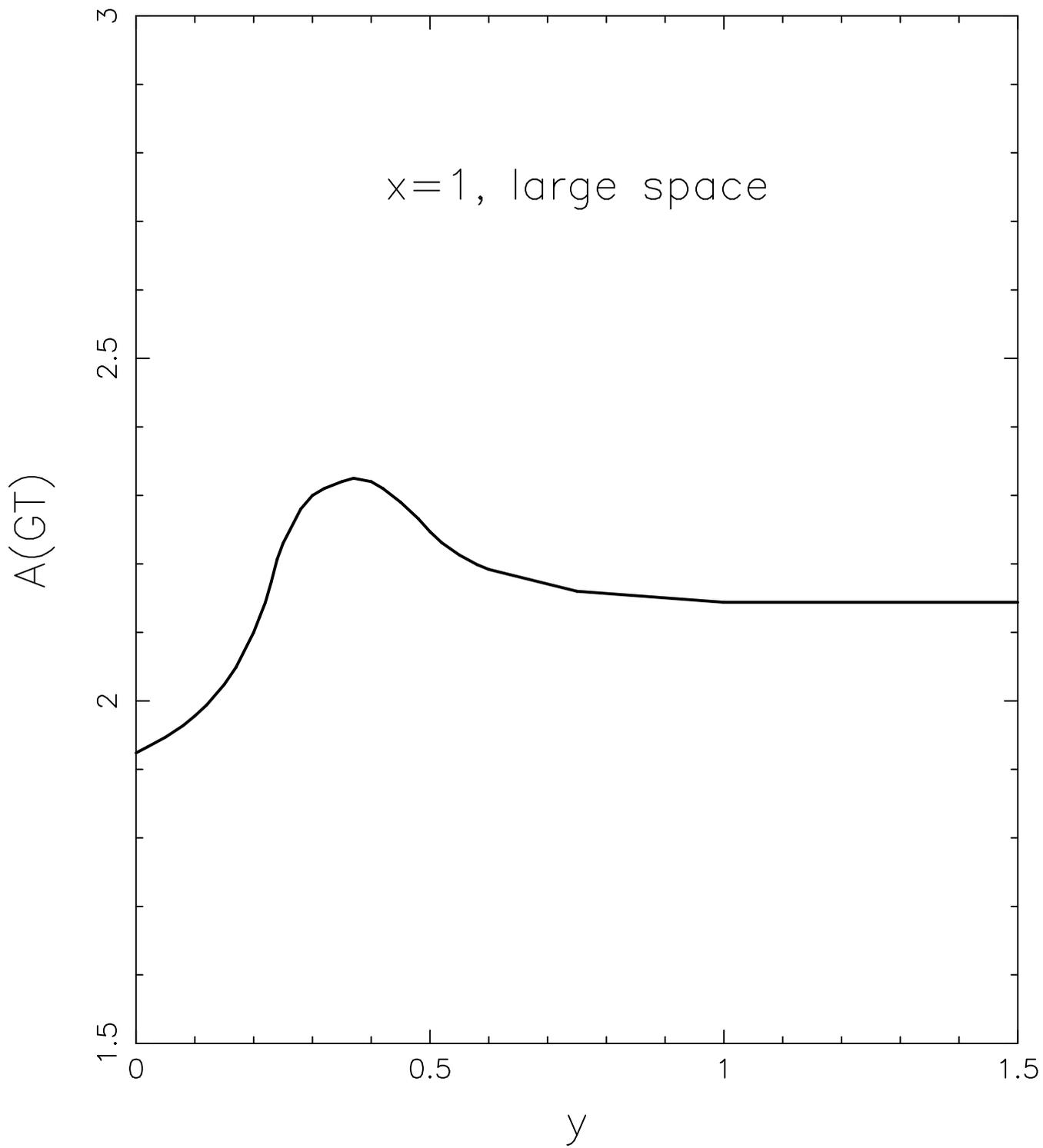
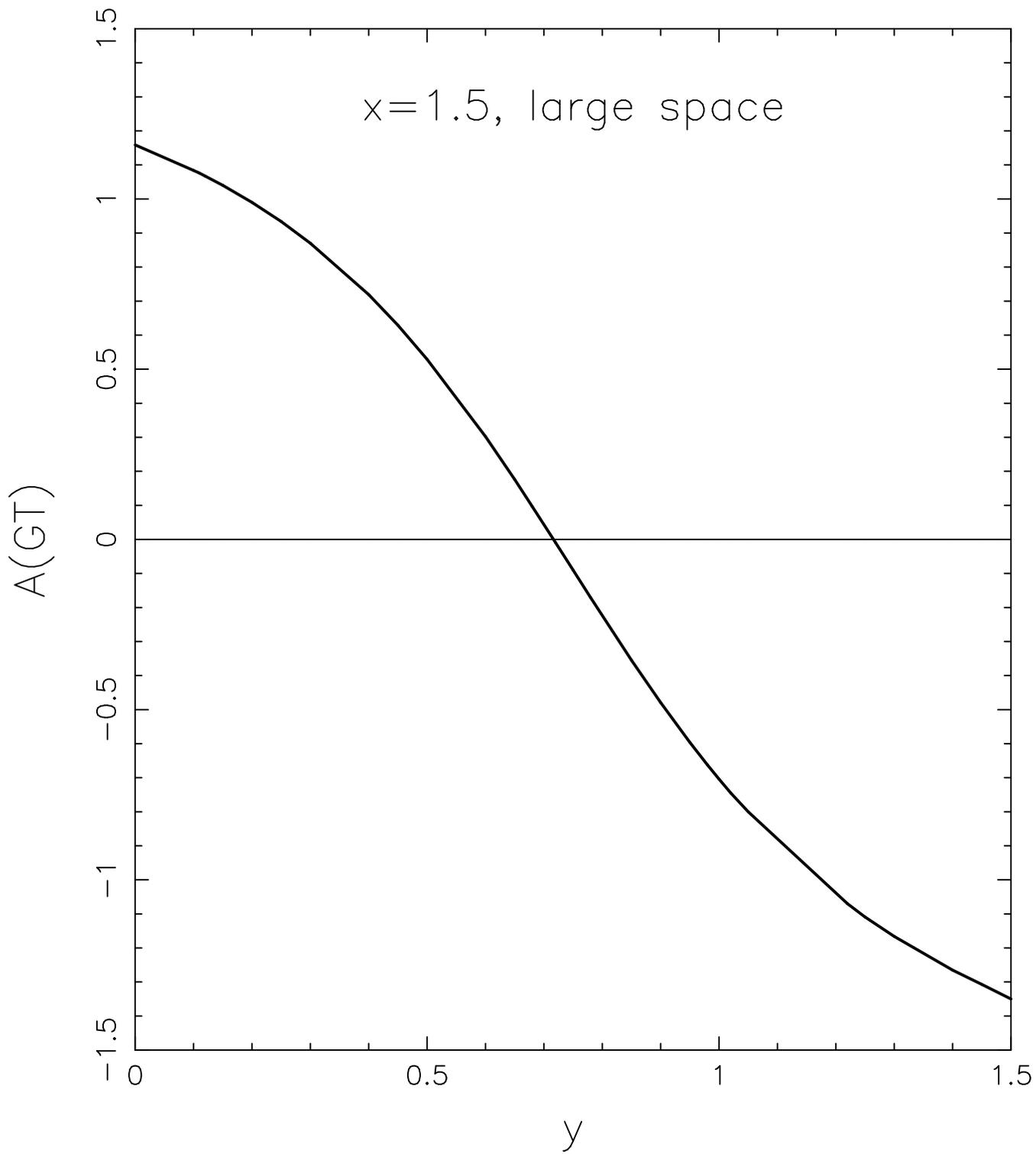
\end{document}